\documentclass[twocolumn,showpacs,preprintnumbers,amsmath,amssymb]{revtex4}
\usepackage{graphicx}
\usepackage{dcolumn}
\usepackage{epsfig}
\usepackage{bm}
\usepackage{bbm}
\def\beq{\begin{equation}}
\def\eeq{\end{equation}}
\def\be{\begin{equation}}
\def\ee{\end{equation}}
\def\bea{\begin{eqnarray}}
\def\eea{\end{eqnarray}}

\begin{document}

\title{Oscillations of high energy neutrinos in matter: Precise formalism 
\\ 
and parametric resonance}

\author{E. Kh. Akhmedov$^{a,b}$, M. Maltoni$^{c}$  and A. Yu. Smirnov$^{d,e}$} 
 \affiliation{
$^a$ Physik Department T30, Technische Universit\"at M\"unchen \\
James-Franck Stra\ss e, D-85748 Garching, Germany\\
$^b$ Kurchatov Institute,  Moscow, Russia\\
$^c$ C.N.~Yang Institute for Theoretical Physics \\
\mbox{SUNY at Stony Brook, Stony Brook, NY 11794-3840, USA} \\ 
$^d$ ICTP, Strada Costiera 11, 34014 Trieste, Italy\\ 
$^e$ Institute for Nuclear Research, Russian Academy of Sciences, Moscow, 
Russia }

\begin{abstract}
We present a formalism for precise description of oscillation phenomena 
in matter at high energies or high densities, $V > \Delta m^2/2E$, where 
$V$ is the matter-induced potential of neutrinos. The accuracy of the
approximation is determined by the quantity $\sin^2 2\theta_m \Delta
V/2\pi V $, where $\theta_m$ is the mixing angle in matter and $\Delta V$
is a typical change of the potential over the oscillation length
($l \sim 2\pi/V$). We derive simple and physically transparent formulas
for the oscillation probabilities, which are valid for arbitrary matter
density profiles. They can be applied to oscillations of high energy 
($E > 10$ GeV) accelerator, atmospheric and cosmic neutrinos in the
matter of the Earth, substantially simplifying numerical calculations 
and providing an insight into the physics of neutrino oscillations in 
matter. The effect of parametric enhancement of the oscillations of high 
energy neutrinos is considered. Future high statistics experiments can 
provide an unambiguous evidence for this effect.
\end{abstract}

 \hspace*{3cm} \preprint{YITP-SB-05-16}

\maketitle

{\it 1. Introduction.} Neutrino physics enters a new phase now, where the 
objectives are precision measurements of the parameters, studies of 
subleading oscillation effects and searches for new physics beyond the already
standard picture, which includes non-zero neutrino masses and mixing. Detection 
of neutrinos from new sources, in particular, of cosmic neutrinos, is in the 
agenda.  

Substantial new information is expected from the studies of high 
energy ($E > 1$ GeV) neutrinos. This includes investigations of atmospheric 
neutrinos with new large volume detectors~\cite{fatm}, long baseline 
accelerator experiments~\cite{LBL} and detection of cosmic neutrinos from 
galactic and extragalactic sources, as well as of neutrinos produced in 
interactions of cosmic rays with the solar atmosphere \cite{solatm}. 
Another possible source of neutrinos is annihilation of hypothetical WIMP's 
in the center of the Earth and the Sun \cite{WIMPS}. 
In all these cases beams of high energy neutrinos can propagate significant  
distances in the matter of the Earth (or of the Sun)  
and therefore undergo oscillations/conversions in matter.  

Increased accuracy and reach of neutrino experiments put forward new 
and more challenging demands to the theoretical description of neutrino 
oscillations. In the present Letter we derive very simple and accurate 
analytic formulas describing oscillations of neutrinos in matter. Our 
primary goal is to study oscillations of high energy neutrinos \cite{note1}, 
but the formulas we obtain are actually applicable in a wide range of neutrino 
energies. They simplify substantially numerical calculations and allow a deep 
insight into the physics of neutrino conversions in matter. In particular, 
they provide a useful tool for studying parametric resonance enhancement 
of neutrino oscillations. 

The parametric resonance can occur in oscillating systems with varying 
parameters due to a specific correlation between the rate of the change of 
these parameters and the values of the parameters themselves. In the case of 
neutrino oscillations in matter, the parametric enhancement is realized when 
the variation of the matter density along the neutrino trajectory is in a 
certain way correlated with the change of the oscillation phase 
\cite{ETC,Akh1}. This effect is different from the MSW effect \cite{W,MS}, 
and yet can result in a strong amplification of the oscillations. We apply 
the formalism developed here to study the parametric resonance in oscillations 
of high energy neutrinos in the Earth. 

{\it 2. Formalism. }
We consider oscillations in the 3-flavor neutrino system ($\nu_e,\,\nu_\mu,\,
\nu_\tau$), with the mass squared differences $\Delta m_{31}^2$ and 
$\Delta m_{21}^2$ responsible for the oscillations of atmospheric and solar 
neutrinos, respectively. In certain situations, the full three-flavor 
problem is approximately reduced to the effective two-flavor ones, in which 
the electron neutrino $\nu_e$ oscillates into a combination of $\nu_\mu$ and 
$\nu_\tau$. We shall be mainly interested in oscillations of neutrinos with 
energies $E > \Delta m_{31}^2/2V$, where the matter-induced potential of 
neutrinos $V(x) \equiv \sqrt{2} G_F N_e(x)$, with $N_e(x)$ the electron 
number density in matter and $G_F$ the Fermi constant. Numerically, this 
corresponds to $E > 8- 10$ GeV for the matter of the Earth. In this case 
the 1-2 mixing is strongly suppressed by matter, and the problem is reduced to 
an effective two-flavor one, described by the potential $V$, mass squared 
difference $\Delta m^2\equiv \Delta m_{31}^2$ and vacuum mixing angle 
$\theta\equiv \theta_{13}$ (which is assumed to be non-zero) \cite{ADLS}. 
In particular, the oscillations of electron neutrinos are determined by the 
transition probability $P_2 \equiv P(\nu_e \leftrightarrow \nu_a)$, where 
$\nu_a = \sin\theta_{23}\nu_{\mu} + \cos\theta_{23}\nu_{\tau}$ is the state 
which mixes with $\nu_e$ in the third mass eigenstate. In terms of $P_2$ the 
flavor transition probabilities are $P(\nu_e \rightarrow \nu_\mu) = P(\nu_\mu 
\rightarrow \nu_e) = \sin^2 \theta_{23} P_2$, $P(\nu_e \rightarrow \nu_\tau) = 
P(\nu_{\tau} \rightarrow  \nu_e) = \cos^2 \theta_{23}P_2$ \cite{ADLS}. The 
third state (orthogonal to $\nu_a$) decouples from the rest of the system and 
evolves independently.  

In the ($\nu_e,\,\nu_a)$ basis, the evolution matrix $S(x)$ describing neutrino 
oscillations satisfies the equation   
\begin{equation}
i \frac{d S}{dx} = H(x) S\,,  
\label{evoleq}
\end{equation}
with the Hamiltonian 
\begin{equation}
\label{H}
H(x) = \frac{V}{2}   
\left( \! \! \begin{array}{cc} 
~1 & ~~0 ~\\
~0 & -1   
\end{array}  \! \! \right)  + 
\delta
\left(  \! \! \begin{array}{cc} 
- \cos 2\theta & ~\sin 2\theta \\
~\sin 2\theta &  ~\cos 2\theta   
\end{array}  \! \! \right).  
\end{equation}
Here  
\begin{equation}
\delta \equiv \Delta m^2/4E\,,
\end{equation}
and the first (potential) term dominates in the high energy limit. However, 
in most situations of interest the neutrino path length in matter $L$ 
satisfies $\delta \cdot L \gtrsim  1$, therefore we cannot consider the 
whole second term as a small perturbation, and the effect of $\delta$ on 
the neutrino energy level splitting should be taken into account. 
For this reason we split the Hamiltonian as  
\begin{equation}
H = H_0 + H_I, 
\label{H0I}
\end{equation}
with  
\begin{equation}
\label{Hsum}
H_0 = \omega   
\left( \! \! \begin{array}{cc} 
~1 & ~~0 ~\\
~0 & -1   
\end{array}  \! \! \right), ~~~ 
H_I = \sin 2\theta \, \delta \, \left(  \! \! \begin{array}{cc} 
- \epsilon & ~1 \\
~1 & ~\epsilon   
\end{array}  \! \! \right)\, .
\end{equation}
Here 
\be
\omega(x) \equiv \sqrt{(V/2 - \delta \cos 2 \theta)^2 + \delta^2 \sin^2 
2\theta}\,,   
\label{split}
\ee
$2\omega$ being the difference of the eigenvalues of $H(x)$;  
\begin{equation}
\label{eps}
\epsilon \equiv \frac{\cos 2\theta\,\delta - V/2 + \omega }
{ \sin 2\theta \, \delta } \approx \frac{\delta }{V} \sin 2\theta \ll 1. 
\end{equation} 

The ratio of the second and the first terms in the Hamiltonian in (\ref{Hsum}) 
is determined by the mixing angle in matter $\theta_m$:
\be
\label{xi}
\frac{\sin 2\theta\, \delta }{\omega}= \sin 2\theta_{m}\,.
\ee
Therefore for $\sin 2\theta_{m} \ll 1$ the term   
$H_{I}$  can be considered as a perturbation. 
Furthermore, according to (\ref{eps}), $\epsilon \sim \sin 2\theta_m$, so 
that the diagonal terms in $H_I$ can be neglected in the lowest approximation. 

We seek the solution of eq. (\ref{evoleq}) in the form 
\begin{equation}
\label{Smatr} 
S = S_0 \cdot S_I\,, 
\end{equation}
where $S_0$  is the solution of the evolution equation with $H$ replaced by  
$H_0$.
{}From (\ref{Hsum}) we find 
\begin{equation}
\label{Szero}
S_0(x) =  
\left(
\begin{array}{cc} 
e^{- i \phi(x)}~  & 0~ \\
      0                        &    e^{i \phi (x)}~
\end{array} 
\right), 
\end{equation} 
where    
\be 
\phi(x) \equiv  \int_{0}^{x} dx' \, \omega(x')
\label{phim}
\ee
is the adiabatic phase. 
Then, according to (\ref{evoleq}), (\ref{H0I}) and (\ref{Smatr}), the 
matrix $S_{I}$ satisfies the equation 
\be
\label{eqSdelta}
i \frac{d S_{I}}{dx} = S_0^{-1} H_{I} S_0 \, S_{I} = \tilde{H}_I 
S_{I}\,,
\ee
where $\tilde{H}_I \equiv S_0^{-1} H_{I} S_0$ is the perturbation Hamiltonian 
in the ``interaction'' representation. Eq. (\ref{eqSdelta}) can be solved by 
iterations: $S_I = \mathbbm{1} + S_{I}^{(1)} + ...$,  which leads to the 
standard perturbation series for the $S$ matrix.  For neutrino propagation 
between  $x = 0$ and $x= L$ we have  
\be
S_{I}(L)=\mathbbm{1}-i\int_0^L dx \tilde{H}_I(x)-\int_0^L dx \tilde{H}_I(x) 
\int_0^x dx' \tilde{H}_I(x') ...
\ee
Taking $S_I$ to the first order, we obtain from (\ref{Smatr}) the evolution 
matrix 
\be
\label{A}
S(L) =  S_0(L) 
\left[\mathbbm{1}  - i \delta \sin 2\theta  \int_0^L dx    
\left(\begin{array}{cc} 
  0 &  e^{i 2 \phi(x)}~\\
e^{-i 2 \phi (x)}  &   0~   
\end{array} 
\right)
\right].   
\ee
The  $\nu_{e} \leftrightarrow \nu_a$ transition probability $P_2(L)$ 
is given by the squared modulus of the off-diagonal element $[S(L)]_{a e}$:  
\be
\label{prob}
P_2 = \delta^2 \sin^2 2\theta \left| \int_0^L dx  ~e^{-i 2 \phi(x)} 
\right|^2. 
\ee

For density profiles that are symmetric with respect to the center of the 
neutrino trajectory, $V(x) = V(L - x)$, eq. (\ref{prob}) gives 
\be
\label{probsym}
P_2 = 4\left(\frac{\Delta m^2 }{4E}\right)^2 
\sin^2 2\theta \left[ \int_0^{L/2}dz \cos 2\phi(z)\right]^2, 
\ee
where $z = x - L/2$ is the distance from the midpoint of the trajectory 
and $\phi(z)$ is the phase acquired between this midpoint and the point $z$. 

The transition probability $P_2$ scales with neutrino energy essentially
as $E^{-2}$. The accuracy of eq. (\ref{prob}) also improves with energy as
$E^{-2}$.  This is illustrated by panels (a) and (b) of fig.~1, which show
$P_2$ as a function of neutrino energy for several trajectories inside the
Earth and as a function of 
the zenith angle
of the neutrino trajectory $\Theta$
for several neutrino energies. One can see that already
for $E\gtrsim 8$ MeV the accuracy of our analytic formula is extremely 
good.
Note that when neutrinos do not cross the Earth's core ($\cos\Theta>
-0.837$) and so experience a slowly changing potential $V(x)$, the 
accuracy of
the approximation (\ref{prob}) is very good even in the MSW resonance 
region
$E\sim $5 -- 8 GeV. The accuracy of eq.~(\ref{prob}) is also good for
energies below $\sim 2$ GeV (not shown in the figure); however, in this 
region
the domain of the applicability of (\ref{prob}) is relatively narrow,
since for $E\lesssim 0.5$ GeV the oscillations driven by the ``solar''
parameters ($\Delta m_{21}^2,\,\theta_{12}$) can no longer be neglected. 

To understand the remarkable accuracy of eq.~(\ref{prob}), we find the
correction $\Delta P_2$ to the transition probability in (\ref{prob}) 
emerging in the next nontrivial order in $H_I$. Note that from the above 
considerations one can expect $\Delta P_2/P_2$ to be proportional to 
$\sin 2\theta_m$. Furthermore, as can be easily seen, for uniform matter 
eq.~(\ref{prob}) reproduces  the exact transition probability; therefore one 
expects $\Delta P_2/P_2\propto V'$. A straightforward calculation indeed gives 
\be
\frac{\Delta P_2}{P_2}\simeq \sin^2 2\theta_m\,\frac{\Delta V}{4 \pi \omega} 
\simeq \sin^2 2\theta_m \, \frac{\Delta V}{2\pi V}\,,
\ee
where $\Delta V$ is the change of the potential over the oscillation length 
$\pi/\omega$, and the last equality holds in the high energy regime. For 
slowly changing density this is equivalent to 
\be
\frac{\Delta P_2}{P_2}\simeq \sin^2 2\theta_m \,\frac{V'}{4 \omega^2}\,.
\ee
Introducing the adiabaticity parameter $\gamma = 4\pi\omega/(\sin 
2\theta_m \Delta V)$, we find that $\Delta P_2/P_2 \simeq \sin 2\theta_m 
\gamma^{-1}$, and therefore for small mixing in matter our approximation 
is better than the adiabatic one. At the same time, for $\Delta V/4\pi\omega 
<1$ it is better than the simple expansion in powers of $\sin^2 2\theta_m$.

The matter density profile of the Earth satisfies $V'/ V^2 \lesssim
0.5$, and therefore for neutrino oscillations in the Earth our approximation
is expected to work well when $\sin^2 2\theta_m\ll 1$. This is fulfilled
in the high energy (or, equivalently, high density) limit $E V/\Delta
m^2 \gg \cos 2\theta$, i.e. above the MSW resonance. If the vacuum
mixing angle is small (i.e. $\theta=\theta_{13}$), our expansion parameter
is also small below the resonance. The above formalism applies 
in this low energy case as well, with only minor modifications: the sign of 
$H_0$ in (\ref{Hsum}) has to be flipped, and correspondingly one has to 
replace $\omega\to -\omega$ in eq. (\ref{eps}). These modifications are 
necessary because of the interchange of the eigenvalues of the Hamiltonian 
upon crossing the MSW resonance. Expressions for the transition  probability 
in eqs.~(\ref{prob}), (\ref{probsym}) remain unchanged. Thus, our results 
are in general valid outside the MSW resonance region, which for small 
$\theta$ is very narrow. For the non-resonant channels ($\bar{\nu}$ channels 
for $\Delta m^2>0$ or $\nu$ channels for $\Delta m^2<0$) and small vacuum 
mixing  our formulas are valid in the whole diapason of energies because 
$\sin 2\theta_m$ is always small.

If $\theta_{13}$ is very small or vanishes, $\nu_e\leftrightarrow 
\nu_{\mu,\tau}$ oscillations are driven by $\Delta m^2=\Delta m_{21}^2$ 
and the large mixing angle $\theta=\theta_{12}$.  The oscillation 
probabilities can then be expressed through another effective two-flavor 
probability, $\tilde{P}_2 \equiv \tilde{P}_2(\Delta m_{21}^2,\,\theta_{12},
\,V(x))$, in terms of which the flavor transition probabilities are $P(\nu_e 
\leftrightarrow \nu_\mu) = \cos^2 \theta_{23}\tilde{P}_2$, $P(\nu_e 
\leftrightarrow \nu_\tau) =  \sin^2 \theta_{23}\tilde{P}_2$ \cite{PerSm}. 
For $E V/\Delta m_{21}^2 \gg \cos 2\theta_{12}$ (which for the typical 
densities inside the Earth corresponds to $E\gtrsim 0.5$ GeV), the 
probability $\tilde{P}_2$ is very well approximated by eq.~(\ref{prob}).

Let us consider the case of symmetric matter. Integrating (\ref{probsym})
by parts, one finds
\be 
P_2 = \sin^2 2\theta_m^0 \left[ \sin \phi_L + \omega_0 \int^{L/2}_{0} dz 
\frac{d\omega}{dz} \frac{1}{\omega^2} \sin 2 \phi(z)
\right]^2\,,
\label{diff}
\ee
where $\theta_m^0 \equiv \theta_m (V_0)$, $\omega_0 \equiv \omega (V_0)$,
$V_0$ being the potential 
at the initial and final points of the neutrino trajectory, and $\phi_L$ 
is the adiabatic phase acquired along the entire neutrino path.   
If the potential changes slowly with distance, so that $\omega^{-2}d\omega/dz 
\ll 1$, the second term in (\ref{diff}) can be neglected, and $P_2$ reduces 
to the usual adiabatic probability in symmetric matter: $P_{\rm adiab} = 
\sin^2 2\theta_m^0  \sin^2 \phi_L$. The second term in (\ref{diff}) describes 
the effects of violation of adiabaticity. 

Let us apply the above results to neutrino beams (atmospheric, accelerator, 
cosmic neutrinos) crossing the Earth. According to the PREM model 
\cite{PREM}, the Earth density profile can be described as several spherical 
shells of radii $R_i$ with rather smooth density change within the shells and 
sharp change at the borders between them. Then, along a direction from the
center of the Earth outwards, $\omega(z)$ decreases abruptly from 
$\omega_i^+$ to $\omega_i^-$ in very narrow regions around $R_i$. 
Therefore $d\omega/dz$ is large in these narrow regions and small  
outside them. The integration in (\ref{diff}) can then be easily done, 
leading to  
\be 
P_2 = \sin^2 2\theta_m^0 \left[  \sin \phi_L -  \omega_0 \Sigma_i 
\frac{\omega_i^+ - \omega_i^-}{\omega_i^+ \omega_i^-}
\sin 2\phi_i \right]^2 .
\label{diffint}
\ee
Here $\phi_i$ is the adiabatic phase acquired by neutrinos between the points 
$z=0$ and $z=R_i$. We will use eq.~(\ref{diffint}) for interpreting the 
results of our calculations.  
 
\begin{figure*}
\begin{center} 
\epsfig{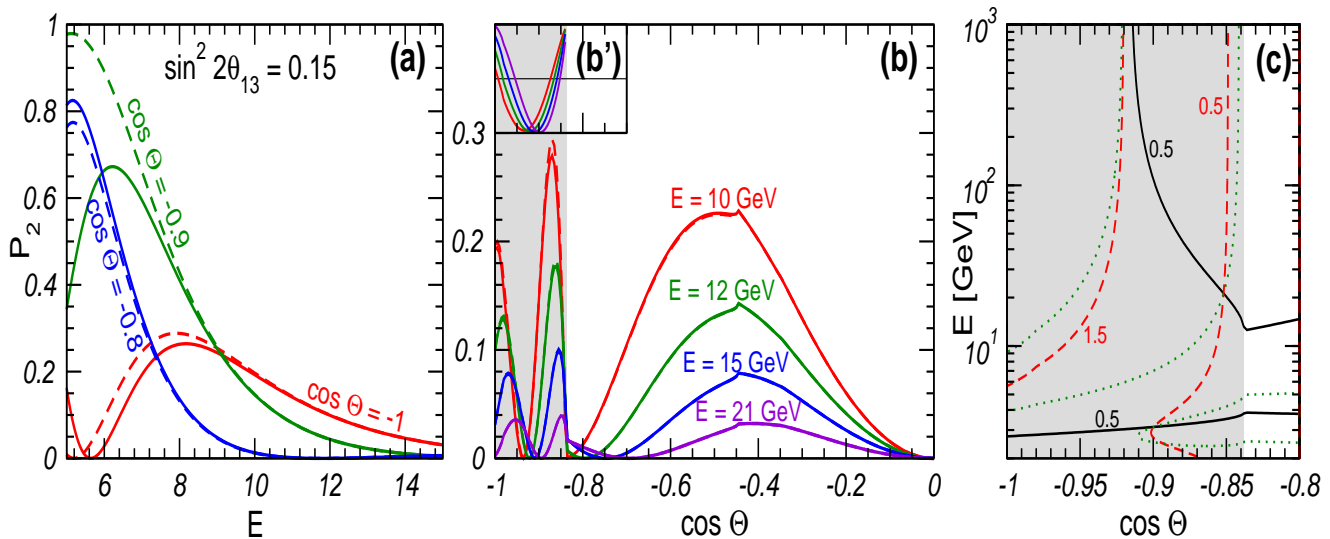}
\caption{\label{fig1} Panels (a) and (b): transition probability $P_2$ 
{\it vs} neutrino energy for different trajectories inside the Earth [panel 
(a)] and 
{\it vs} $\cos\Theta$ for different neutrino energies [panel 
(b)]. Solid curves are the results of exact numerical calculations, dashed 
curves are obtained using formula (\ref{prob}). Small window (b') in panel 
(b) shows the values of the parameter $X_3$ [eq.~(\ref{X3})] calculated in 
the three-layer model of the Earth's density. Panel (c): contours of constant 
phases $\phi_{m}$ (solid curves), $\phi_{c}$ (dashed curves) and of constant 
$X_3=0$ (dotted curves). The numbers at the curves are the values of the 
phases in units of $\pi$. The shaded areas in panesls (b) and (c) 
correspond to the Earth's core. For all panels we take $\sin^2 2\theta_{13} 
= 0.15$ and $\Delta m^2 = 2 \cdot 10^{-3} $ eV$^2$. }
\end{center} 
\end{figure*}

{\it 3. Parametric enhancement of oscillations}. 
In the PREM model of the Earth density all the density jumps between 
different shells except those between the mantle and core are relatively 
small \cite{PREM}. Therefore the density profile felt by neutrinos 
crossing the core of the Earth can be approximated by three layers (mantle 
- core - mantle) with slow density changes within the layers.
Eq. (\ref{diffint}) then gives 
\be 
P_2 = \sin^2 2\theta_m^0 \left[\sin(\phi_c+2\phi_m) -  
\frac{\omega_0}{\omega_m} \left(1-\frac{\omega_m}{\omega_c} \right)
\sin \phi_c \right]^2, 
\label{probpar}
\ee  
where $\omega_m$ and  $\omega_c$ are the values of $\omega(x)$ in the 
mantle and core on the respective sides of their border, and 
$\phi_m$ and $\phi_c$ are the phases acquired in the mantle (one layer) and 
core. Eq. (\ref{probpar}) corresponds to the adiabatic neutrino propagation 
inside the core and mantle and violation of the adiabaticity at the borders 
between them. In the approximation of constant effective densities 
in the mantle and core we have $\omega_0 = \omega_m^{\rm eff}$, and the 
non-adiabatic term is proportional to $(1 - \omega_c^{\rm eff}/
\omega_m^{\rm eff})$. 

For neutrino trajectories that cross the mantle only ($\phi_c = 0$), 
eq.~(\ref{probpar}) reduces to the adiabatic probability. The passage of 
neutrinos through the core can lead to an enhancement of the oscillations. As 
follows from (\ref{probpar}), the maximum enhancement of the probability can 
be achieved when $\sin(\phi_c + 2\phi_m)$ and $\sin\phi_c$ are of opposite 
sign and maximal amplitude: 
\be
\sin \phi_c  = - \sin(\phi_c + 2 \phi_m) =  \pm 1, 
\ee
that is, when  
\be
\label{pcond}
\phi_c  = \pm \frac{\pi}{2} + 2\pi n, ~~~ 
\phi_m = \pm \frac{\pi}{2} +  2\pi k\,. 
\ee
Here $n$ and $k$ are integers, and the signs in front of $\pi/2$ are 
correlated. In this case the enhancement factor is 
\be
\label{ench}
\frac{P_2^{max}}{\sin^2 2\theta_m}  
 = \left[ 1 + \frac{\omega_0}{\omega_m}\left(1-\frac{\omega_m}{\omega_c}
 \right)\right]^2 
\approx \left( 2 - \frac{V_m}{V_c}  \right)^2 \approx 2.5\,,
\ee
where $\sin^2 2\theta_m$ in the denominator corresponds to the maximum 
possible transition probability for neutrinos crossing only the mantle, 
and we have taken into account that at high energies 
$\omega_m/\omega_c\approx V_m/V_c$. 

The condition in eq.~(\ref{pcond}) and the enhancement described by 
eq.~(\ref{ench}) are the particular cases of the parametric resonance 
condition and the parametric enhancement of neutrino oscillations 
\cite{ETC,Akh1}. In \cite{Akh2} it was shown that in the case of matter 
consisting of alternating layers of two different constant densities and (in 
general) different widths the parametric resonance condition is  
\be
\label{X3}
X_3\equiv -(\sin\phi_m \cos\phi_c\cos 2\theta_m+\cos\phi_m \sin\phi_c\cos 
2\theta_c) = 0\,,
\ee
where $\theta_{m,c}$ and $\phi_{m,c}$ are the mixing angles and the 
acquired oscillations phases in the layers $m$ and $c$. This condition 
can also be used as an approximate one when matter density inside the 
layers is not constant but varies sufficiently slowly. For neutrino 
oscillations in the Earth we identify the layers $m$ and $c$ with the 
mantle and core. Since in the energy region $\sin 2\theta\,\delta\ll V$
one has $\theta_m\approx \theta_c\approx \pi/2$, condition (\ref{X3})
reduces to
\be
X_3\simeq \sin(\phi_m+\phi_c)=0\,.
\label{X3a}
\ee
Eq.~(\ref{pcond}) is a particular realization of this condition. In the 
high energy limit the parametric resonance condition (\ref{pcond}) was 
previously considered in the framework of active-sterile atmospheric neutrino 
oscillations in \cite{Liu}. It should be noted that, while the realization 
(\ref{pcond}) of condition (\ref{X3}) corresponds to the maximal possible 
parametric enhancement of oscillations of high energy neutrinos in the Earth, 
a sizable amplification is also possible if the equality $X_3=0$ is realized 
differently, i.e. when the two terms in (\ref{X3}) do not separately vanish 
but cancel each other \cite{note2}. 

The parametric resonance conditions in eq. (\ref{X3}) or (\ref{pcond}) require 
a subtle correlation between the matter density of the Earth, the distances 
that neutrinos travel in the Earth's mantle and core (which are not 
independent), and also in general neutrino energy and oscillation parameters. 
Therefore it is far from obvious that these conditions can actually be 
satisfied. Amazingly, this is indeed the case 
\cite{Liu},\cite{Pet},\cite{Akh2}. 

Our present analysis shows that, for $\nu_e \leftrightarrow \nu_{\mu,\tau}$ 
oscillations of high energy neutrinos in the Earth, there are two regions of 
the zenith angles of neutrino trajectories in which (\ref{X3}) is  
satisfied: $\cos\Theta = -1\div -0.93$ and $\cos\Theta = -0.88\div 
-0.84$. This is illustrated in panel (b) of fig. 1. 
Since at high energies matter suppresses neutrino mixing, one could expect 
that for the trajectories crossing the core ($\cos\Theta < -0.837$), where 
the densities are higher, the probability would be suppressed or at least would 
not change. Instead, we see two prominent peaks there, exceeding maximal 
allowed by the MSW effect value of probability $\sin^2 2\theta_m$ by up 
to a factor of 2. This is the result of the parametric enhancement of 
neutrino oscillations. 
As can be seen from the figure, for core-crossing trajectories the positions  
of the peaks of $P_2$ essentially coincide with zeros of $X_3$. 

For neutrino energies $E\simeq 10$ -- 15 GeV, the oscillation phases 
corresponding to the peak with $\cos\Theta \lesssim -0.93$ (the inner peak) 
are $\phi_m\simeq \pi/4$, $\phi_c\simeq 7\pi/4$, while for the peak with 
$\cos\Theta = -0.88\div -0.84$ (the outer peak), they are $\phi_m\simeq 
0.35\pi$, $\phi_c\simeq 0.65\pi$. In both peaks to a good accuracy 
$\phi_m+\phi_c=n\pi$, so that eq.~(\ref{X3a}) is satisfied. The phases in 
the outer peak are closer to the realization (\ref{pcond}) of the parametric 
resonance condition, and therefore in this peak the parametric enhancement of 
oscillations is closer to the maximal possible one. From panel (c) of fig. 1 
one can see that at $E\simeq 21$ GeV the maximum enhancement condition 
(\ref{pcond}) can be exactly realized in the outer peak. For neutrinos of 
very high energies ($E\gtrsim 100$ GeV), it can be realized nearly exactly in 
the inner peak, whereas the outer peak becomes slimmer and lower, and at 
energies $E\gtrsim 300$ GeV virtually disappears. 

{\it 4. In conclusion} -- 
We have derived very simple and accurate integral formulas describing 
neutrino oscillations in matter with arbitrary density profiles.  
They can be applied to all possible situations where the effective 
mixing in matter is small. 
In particular, our results can be applied to atmospheric, accelerator 
and cosmic neutrinos crossing the Earth as well as to cosmic neutrinos 
crossing the Sun. They can be used for probing the effects of small jumps in 
the density profile of the Earth as well as evaluating the effects of 
uncertainties in this profile on the interpretation of the results of 
future  accelerator experiments. They can also be employed for studying 
the effects of proper averaging over the energy spectrum of the neutrino 
beam. We used the obtained formulas to study the parametric enhancement of 
oscillations of high energy neutrinos in the Earth and identified two peaks 
in the zenith angle distribution of core-crossing neutrinos, which are due to 
the parametric effects. 
Observation of these peaks in future high statistics experiments with high 
energy neutrinos would provide an unambiguous evidence for the parametric 
resonance effects in neutrino oscillations in matter. 

This work was supported in part by SFB-375 f\"ur Astro-Teilchenphysik der 
Deutschen Forschungsgemeinschaft (EA), the National Science Foundation 
grant PHY0354776 (MM) and by the Alexander von Humboldt Foundation (AS).



\begin{thebibliography}{99}
\bibitem{note2} In principle, the parametric resonance can also lead to a 
parametric suppression of oscillations; however, this does not happen for 
oscillations of high energy neutrinos in the Earth.
\end{thebibliography}
\end{document}